\documentclass[%
    reprint,
    superscriptaddress,
    amsmath,amssymb,
    aps,
    prd,
]{revtex4-2}

\usepackage{xcolor}
\usepackage{graphicx}
\usepackage{graphicx}
\usepackage{dcolumn}
\usepackage{bm}
\usepackage[utf8]{inputenc}  
\usepackage[T1]{fontenc}

\begin{document}

    \preprint{APS/123-QED}
    
    \title{Topological Momentum Skyrmions in Mie Scattering Fields}

    \author{Peiyang Chen}
    \affiliation{Centre for Disruptive Photonic Technologies, School of Physical and Mathematical Sciences, Nanyang Technological University, Singapore 637371, Singapore.}
    \affiliation{School of Physics and Astronomy, Shanghai Jiao Tong University, Shanghai, 200240, China.}
    \affiliation{Zhiyuan College, Shanghai Jiao Tong University, Shanghai, 200240, China.}

    \author{Kai Xiang Lee}
    \affiliation{Centre for Disruptive Photonic Technologies, School of Physical and Mathematical Sciences, Nanyang Technological University, Singapore 637371, Singapore.}
    \affiliation{Centre for Quantum Technologies, National University of Singapore, Singapore 117543, Singapore.}
    
    \author{Tim Colin Meiler}
    \affiliation{Centre for Disruptive Photonic Technologies, School of Physical and Mathematical Sciences, Nanyang Technological University, Singapore 637371, Singapore.}
    \affiliation{Institute of Materials Research and Engineering, A*STAR (Agency for Science Technology and Research), Singapore 138634, Singapore}
    
    \author{Yijie Shen}
    \thanks{yijie.shen@ntu.edu.sg}
    \affiliation{Centre for Disruptive Photonic Technologies, School of Physical and Mathematical Sciences, Nanyang Technological University, Singapore 637371, Singapore.}
    \affiliation{School of Electrical and Electronic Engineering, Nanyang Technological University, Singapore 639798, Singapore.}
    \begin{abstract}
        Topological quasiparticles such as skyrmions and merons have recently attracted enormous attentions in the form of diverse optical degrees of freedom. However, these structures have not been explored in the fundamental momentum vectors of optical fields yet. Here, we reveal the universality of forming skyrmion and meron topological textures from the Poynting vector, canonical momentum, and optical spin field, which are generated from multipole Mie scattering fields. Moreover, we analyze the unconditional topological stability of the skyrmionic momentum fields against perturbation and geometric defects. This work reveals the topological properties of multipole scattered field and will spur new phenomena related to optical forces, metamaterial design and unique light-matter interaction.
    \end{abstract}
    \maketitle

\section{Introduction}
Skyrmions are topological nontrivial vector textures, first proposed in particle physics as a low-energy nucleon model \cite{skyrme1962unified}. Over subsequent decades, interest in skyrmions have translated to the condensed matter domain, and these structures have been realized in a diverse range of systems, including atomic condensates \cite{al2001skyrmions,ruostekoski2001creating}, liquid crystals \cite{fukuda2011quasi,duzgun2021skyrmion,tai2024field}, and chiral magnets \cite{muhlbauer2009skyrmion,nagaosa2013topological,bogdanov2020physical}. In particular, skyrmions have demonstrated their potential as novel topologically-protected information carriers for local large-density data storage~\cite{fert2017magnetic,han2022high,chen2024all}. More recently, optical vector fields have emerged as a rich and versatile platform for studying skyrmionic structures as an analog to skyrmions realized on matter-based fields \cite{shen2024optical}. Dubbed optical skyrmions, these topological quasiparticles of light have realized in various vector fields associated with electromagnetic waves, such as electric and magnetic fields~\cite{tsesses2018optical,davis2020ultrafast,shen2021supertoroidal,wang2024observation,shen2024nondiffracting}, optical spins~\cite{du2019deep,dai2020plasmonic,guo2020meron,lin2022photonic}, polarization Stokes vectors~\cite{gao2020paraxial,shen2021generation,sugic2021particle}, and Poynting vector fields~\cite{wang2024topological}. However, one fundamental vector quantity of the electromagnetic field have thus far been overlooked -- momentum. 

The momentum of light determines the energy transfer of the optical field as it propagates over free space and when it interacts with matter \cite{Stratton_2015}. However, one must be careful in determination of this particular quantity. One common method to quantify the energy flow (and thus momentum) of an optical field is the Poynting vector \cite{poynting1884xv}.  This quantity is a convenient measure, requiring knowledge of the electric and magnetic fields to compute. However, the identification of this quantity with energy flow is not universal, with many irregularities and controversies raised since its conceptualization~\cite{gough1982poynting,minkowski1910grundgleichungen,ibrahim1909elektrodynamik,abraham1910minkowski}. 

In light-matter interactions, optical forces can arise from different physical phenomena. Of particular interest is the canonical momentum density component of the energy flow \cite{ghosh2024canonical}. The canonical momentum is collinear with the phase gradients of the optical field and is the dominant component of radiation pressure force. This component provides a clean and direct approach for measurement and quantification of energetic structures in optical fields \cite{baxter1993canonical} -- with sufficiently small test particles or atoms, one can observe energetic structures at the sub-wavelength scale, below the diffraction limit \cite{afanasev2022superkicks,barnett2013superweak}. 

Poynting vector skyrmions have only recently been discovered in free space, by focusing a pair of counter-propagating cylindrical vector vortex beams through a 4$\pi$ microscopic configuration \cite{wang2024topological}. While similar to the canonical momentum field -- and in some cases, collinear -- the Poynting momentum is an aggregate quantity. 

In this letter, we theoretically demonstrate the formation of optical skyrmions in the momentum vectors of a multipole scattering field. We show that the helicity can be tuned by adjusting the phase difference of multipole sources. Additionally, we demonstrate their topological stability against shifts in the scattering sources.

\begin{figure*}[htp!]
    \includegraphics[width=1\textwidth]{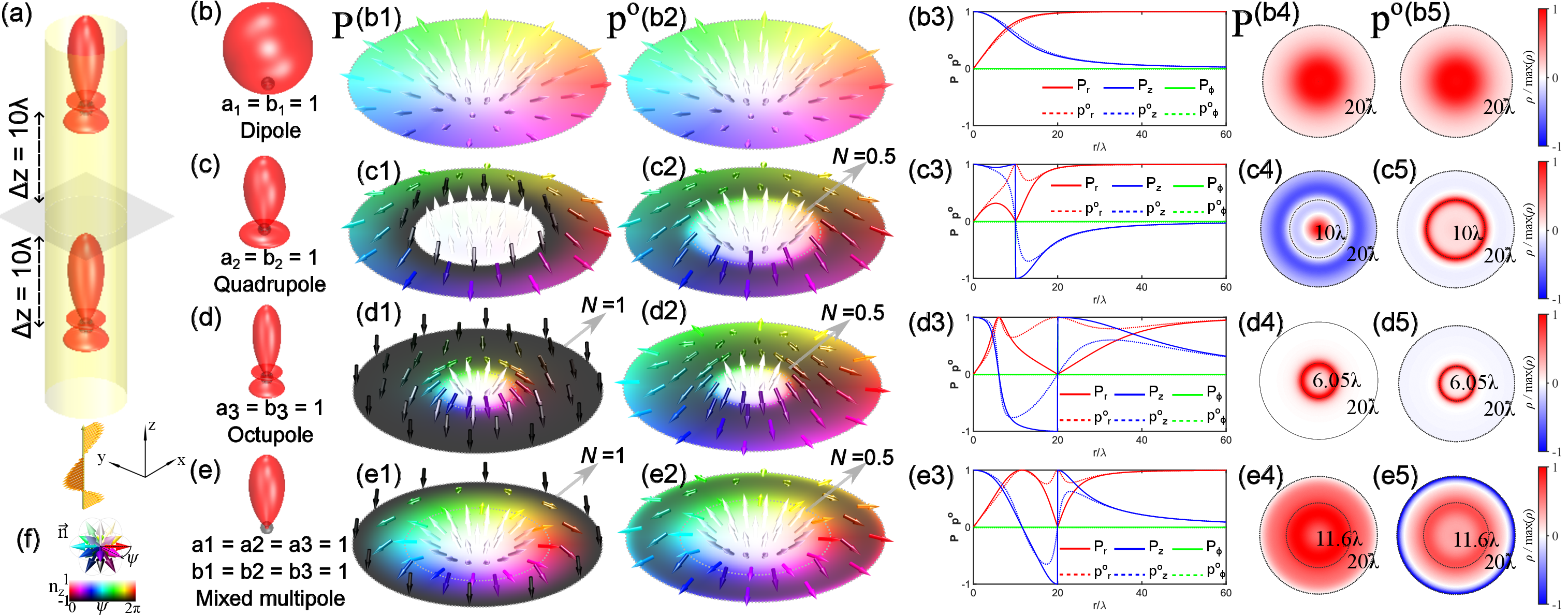}
    \caption{\textbf{Momentum textures of scattered fields from two multipoles.}
    \textbf{(a)} Schematic of the system. A light wave comes in from the bottom and is scattered by two particles symmetrically arranged around the examined plane.
    \textbf{(b-e)} Primary intensity lobes of the scattered radiation fields from multipoles.
    \textbf{(b1-e1)} Normalized Poynting momentum $\mathbf{P}$ textures host skyrmions with topological charge $N=1$ for octupole (d1) and mixed multipole (e1).
    \textbf{(b2-e2)} Normalized canonical momentum $\mathbf{p^o}$ texture shows merons with $N=0.5$ for quadrupole (c2), octupole (d2) and mixed multipole (e2).
    The radius of all these texture plots is $2 \Delta z=20\lambda$.
    \textbf{(f)} Two colorbars indicate the direction of of normalized vectors $\vec{\mathbf{n}}$, out-of plane component $n_z$ and azimuth angle $\psi=\arctan{\frac{n_y}{n_x}}$. This colorbar is consistently used throughout the article to illustrate topological textures.
    \textbf{(b3-e3)}The change of normalized $\mathbf{p}$ and $\mathbf{p^o}$ with respect to $r$, when the multipole source corresponds to \textbf{(b-e)}, respectively.
    \textbf{(b4-e4)}Normalized topological charge density corresponding to (b1-e1)
    \textbf{(b5-e5)}Normalized topological charge density corresponding to  (b2-e2)
    }
    \label{Fig.1} 
\end{figure*}

\section{Momentum Textures of Scattered Fields from Two Multipoles}

We begin by defining key terms. The Poynting vector $\mathbf{P}$ and kinetic momentum density $\mathbf{p}$ are related in the following way:
\begin{align}
    \mathbf{p}=
    \frac{1}{c^2}\mathbf{P}
     =
    \frac{1}{2c^2} \mathfrak{R}(\mathbf{E^*} \times \mathbf{H})
    =
    \mathbf{p^o}+\mathbf{p^s}
\end{align}
where $c$ is the speed of light, $\mathbf{E}$ and $\mathbf{H}$ are electric and magnetic field, respectively.
They can be separated in two components, namely, $\mathbf{p^s}$ arises from the linear momentum analog to the spin angular momentum density \cite{yang2022quantum} and $\mathbf{p^o}$ is the canonical momentum density defined as:
\begin{align}
    \mathbf{p^o}
    =
    \frac{1}{4 \omega} \frak{I}[\varepsilon \mathbf{E^*} \cdot (\nabla) \mathbf{E} + \mu \mathbf{H^*} \cdot (\nabla) \mathbf{H}]
    \label{eq:canonical_momentum_density}
\end{align}
where $\omega$ is the angular frequency, $\varepsilon$ and $\mu$ represent dielectric and magnetic permeability of the propagation medium, respectively. Furthermore, the canonical momentum density $\mathbf{p^o}$ is related to the orbital angular momentum density $\mathbf{L}=\mathbf{r}\times\mathbf{p^o}$, while the spin part of the momentum density $\mathbf{p^s} = \frac{1}{2} \nabla \times \mathbf{S}$ are defined with respect with the spin angular momentum densities $\mathbf{S} = \mathbf{r}\times\mathbf{p^s} = \frac{1}{4\omega} \frak{I} [\varepsilon (\mathbf{E^*} \times \mathbf{E}) + \mu (\mathbf{H^*} \times \mathbf{H})]$. 

A topological configuration can be characterized by its topological charge $N$, 
\begin{align}
    N
    &=
    \frac{1}{4\pi} 
    \iint_{\sigma} 
        \mathbf{n} \cdot \left(
            \frac{\partial \mathbf{n}}{\partial x} \times \frac{\partial \mathbf{n}}{\partial y}
        \right) 
    {\rm d}x {\rm d}y 
    \nonumber
    \\
    &= 
    \iint_{\sigma} \rho(x,y)\ {\rm d}x {\rm d}y
    \label{eq:topological_charge}
 \end{align}
 
where $\rho(x,y)$ represents the topological charge density of the vector field and $\sigma$ traces the boundary of integration. A skyrmion has a topological charge of $N = 1$ while a meron carries a fractional topologial charge of $N=0.5$.

Engineering skyrmions formed by momentum requires consideration of the symmetries within the physical domain. The key factor to constructing the skyrmionic configurations lies in breaking the symmetry in normal direction. In our study, we utilize the asymmetry of the far-field multipole radiation from Mie scattering.

The Mie solution to Maxwell's equations describes the scattering of an electromagnetic plane wave by a spherical particle. The contributions of the different scattered multipoles can be tuned precisely by engineering the properties of the particle \cite{yang2020mie,sugimoto2021colloidal}.

Consider a two-particle system uniformly illuminated by an incoming plane-wave along the inter-particle axis, shown in FIG. \ref{Fig.1}(a). The particles are identical, and have identical scattering properties. We examine the scattered kinetic and canonical momenta in the plane equidistant from both particles. The particles are separated by $2 \Delta z = 20\lambda$. The exact distance between the two scattering particles is irrelevant as long as the scattered radiation is in the far-field, to avoid any complications with evanescent components, which is a key assumption in the multipole expansion.

Mie scattering is a common approach for describing nanoparticle scattering and represents a simplified case of multipole scattering. The Mie coefficients $a_n$ and $b_n$ are corresponding to different orders of multipole components. For expample, $a_1$ is corresponding to electric dipole and $b_3$ is corresponding to magnetic octupole. To tune Mie coefficients or different multipole components, a well-established approach is to construct a metasuface consisting of a periodic array of nanoparticles of identical shape with inversion symmetry, e.g. sphere, cylinder, cube \cite{allayarov2024multiresonant}. Taking an array of cone-shaped nanoparticles as an example, it is possible to obtain multipole components with different proportions and phases by tuning the particle size and the wavelength of the incident light \cite{terekhov2019multipole}. By utilizing these degrees of freedom, one can tune the structure of the scattered radiation field to create skyrmions and merons in the far field.

As forward scattering is stronger than backscattering, the overlap in the multipole radiation in the examined plane will be inhomogeneous, and different topological structures can be generated. We examine the symmetric case where the electric and magnetic field components have the same Mie scattering coefficients, which is known as Kerker condition \cite{kerker1983electromagnetic}. We start with three simple yet representative cases: dipole, quadrupole and octupole.

When multipole sources are pure dipoles ($a_1=b_1=1$), merons are formed in the kinetic and canonical momentum fields in the examined plane (FIG. \ref{Fig.1}(b1,b2)). In this case, the meron boundary extends to infinity where their $z$-components vanishes (FIG. \ref{Fig.1}(b3)).

For pure quadrupole sources ($a_2=b_2=1$), a meron is realized in the canonical momentum field within a finite boundary $r=10\lambda$. In contrast, the kinetic momentum field has a discontinuity at $r = 10 \lambda$ (FIG. \ref{Fig.1}(c3)) due to vanishing electric and magnetic fields and a topological invariant cannot be quantified. Similar meron textures appear in the pure even-ordered multipoles (quadrupoles, hexadecapoles, etc.).

In pure octupolar radiation ($a_3=b_3=1$),  the kinetic momentum field hosts a a skyrmion (FIG. \ref{Fig.1}(d1)), whereas  a meron is present in the canonical momentum field (FIG. \ref{Fig.1}(d2)). The boundary of the skyrmion is realized at the radius where the Poynting vector field vanishes, at $r = 20\lambda$ (FIG. \ref{Fig.1}(d3)). Likewise, skyrmions and merons are realized for the pure odd-ordered multipoles except dipoles, e.g. octupoles, dotriacontapoles, and higher orders.

Here, we explain the reasons for the formation of different topological textures in the kinetic momentum and canonical momentum fields. The kinetic momentum field is proportional to the Poynting vector, and we can discuss the appearance of the topological structures in the same context. At the examined plane, the Poynting vector texture arises from the superposition of the polar distribution of intensity lobes of the multipole scatterers. In particular, even-order pure multipolar sources have an equal number of lobes in forward and backward radiation. This allows the Poynting vector field to undergo discontinuity radially before forming a complete skyrmion, where the electromagnetic field vanishes. The odd-order pure multipolar sources have different number of lobes in forward and backward radiation. The asymmetric superposition of forward and backward scattering induces the formation of a skyrmion in Poynting vector field at the center.

The canonical momentum is directed along the phase gradients of the optical field,  so it does not vanish even as the electric and magnetic field vanishes. As such, it does not suffer from the same discontinuities as the Poynting vector, and can maintain a continuous deformation to infinity which can generate topological structures.

Lastly, let us consider a superposition of multipoles with the same amplitude in order to determine who dominates in mixed multipole. We choose octupole as the highest order multipole component in mixed multipole since octupole source is the simplest source to generate $\mathbf{P}$ skyrmion and $\mathbf{p^o}$ meron in our system. We observe the formation of a skyrmion in the kinetic momentum field within a radius of $20\lambda$ (FIG. \ref{Fig.1}(e1)), and a meron in the canonical momentum (FIG. \ref{Fig.1}(e2)). In fact, the highest order of mixed multipole sources with equal weights determines the features of the momentum field, including $\mathbf{P}$ field and $\mathbf{p^o}$ field, such as the number of reversals in the $z$ component of the momentum.

\begin{figure}
    \includegraphics[width=0.49\textwidth]{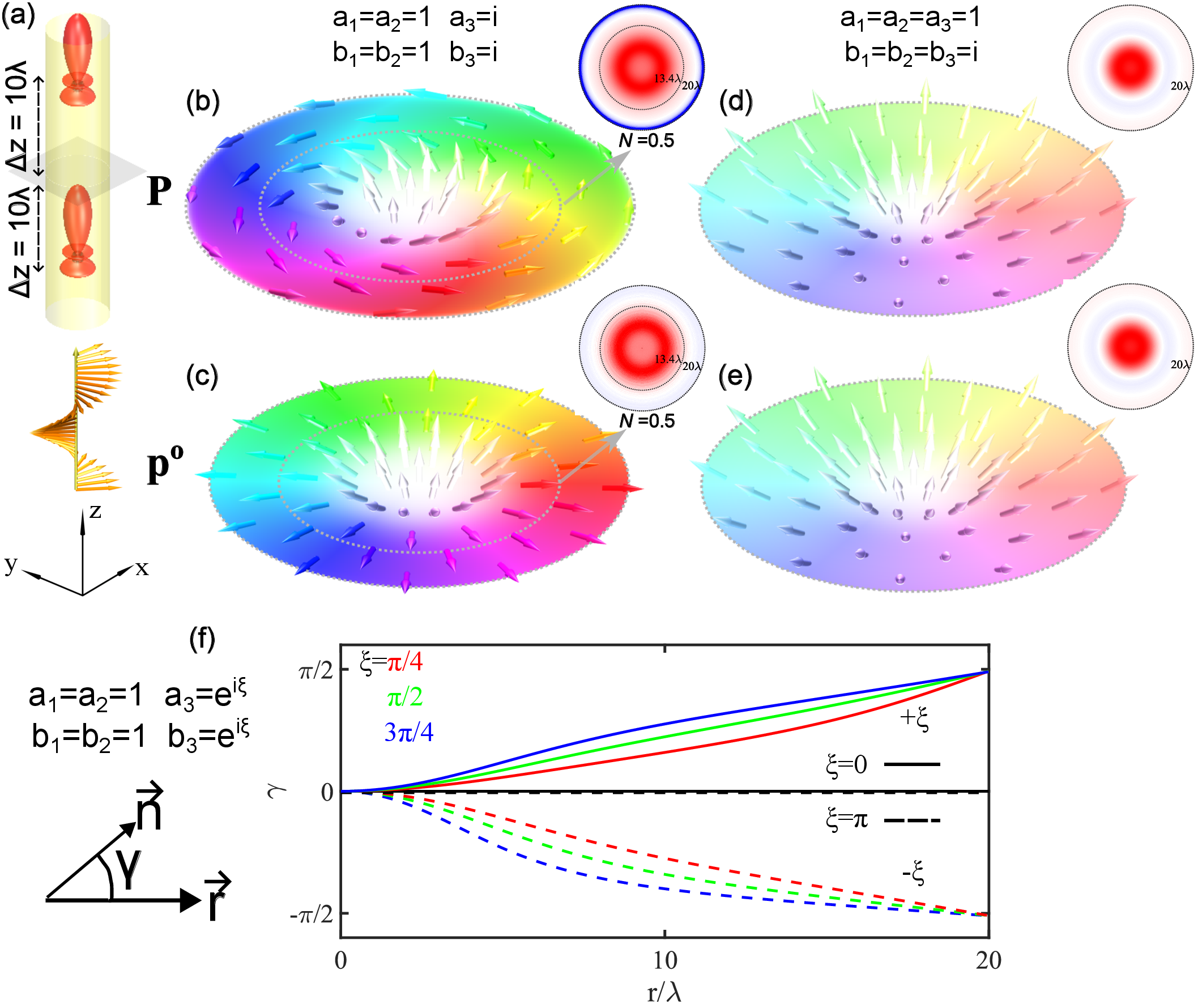}
    \caption{\textbf{The tunability of the helicity of the topological structure in momentum field.} \textbf{(a)} Schematic diagram of the two-particle system, similar to FIG. \ref{Fig.1}. \textbf{(b-e)} The kinetic and canonical momentum distributions respectively, when (b-c) $a_{1,2}=b_{1,2}=1$ and $a_3=b_3=i$, and (d-e) $a_{1,2,3}=1$ and $b_{1,2,3}=i$. \textbf{(f)} The change of $\gamma$ with respect to $r$ for different multipole source with phase difference. $\gamma \in [-\pi, \pi]$ represents the angle between the in-plane field vector(Poynting vector here) and the radial position vector }
    \label{Fig.3}
\end{figure}

\section{Tuning of helicity}

Quasiparticle textures can also be charactized by their helicity $\gamma \in [-\pi, \pi]$, which preserves the topological invariants under a global azimuthal rotation, that yields the well-known Néel and Bloch-type skyrmions \cite{zhang2021bloch}. To introduce helicity, one must identify a global $U(1)$ degree-of-freedom to control. In our case, we have two ways of control -- the phase of the Mie coefficients, and the ellipticity of the irradiation field.

While keeping the multipole coefficients $a_{1,2}=b_{1,2}=1$, we induce a phase to the octupole coefficient $a_3=b_3=i$. Additionally, the incident light beam is circularly polarized in order to couple to both in-phase and quadrature components of the Mie scattering coefficients (FIG. \ref{Fig.3}(a)). The resulting multipole modes are similar to those shown in FIG. \ref{Fig.1}(e), with the octupole component out-of-phase.

The resulting scattered fields are shown in  FIG. \ref{Fig.3}(b,c). We notice that the helicity of the kinetic momentum has been modified, which arises from the phase difference of the highest-order multipole and circularly polarized, incident light changing the polarization of the field at the examined plane. As the polarization of the field does not modify the phase gradient, the canonical momentum remains no helicity. Besides the typical example above, adjusting the phase difference between Mie coefficients enables control over the helicity of the Poynting vector's topological structure (FIG. \ref{Fig.3}(f)).

If we introduce phase differences between electric and magnetic multipole components, such as $a_{1,2,3}=1, \quad b_{1,2,3}=i$, the corresponding Poynting vector texture is shown as FIG. \ref{Fig.3}(d) and canonical momentum texture is shown as FIG. \ref{Fig.3}(e). Both are merons whose boundaries are infinity and exhibit no helicity.

\begin{figure}
    \includegraphics[width=0.4\textwidth]{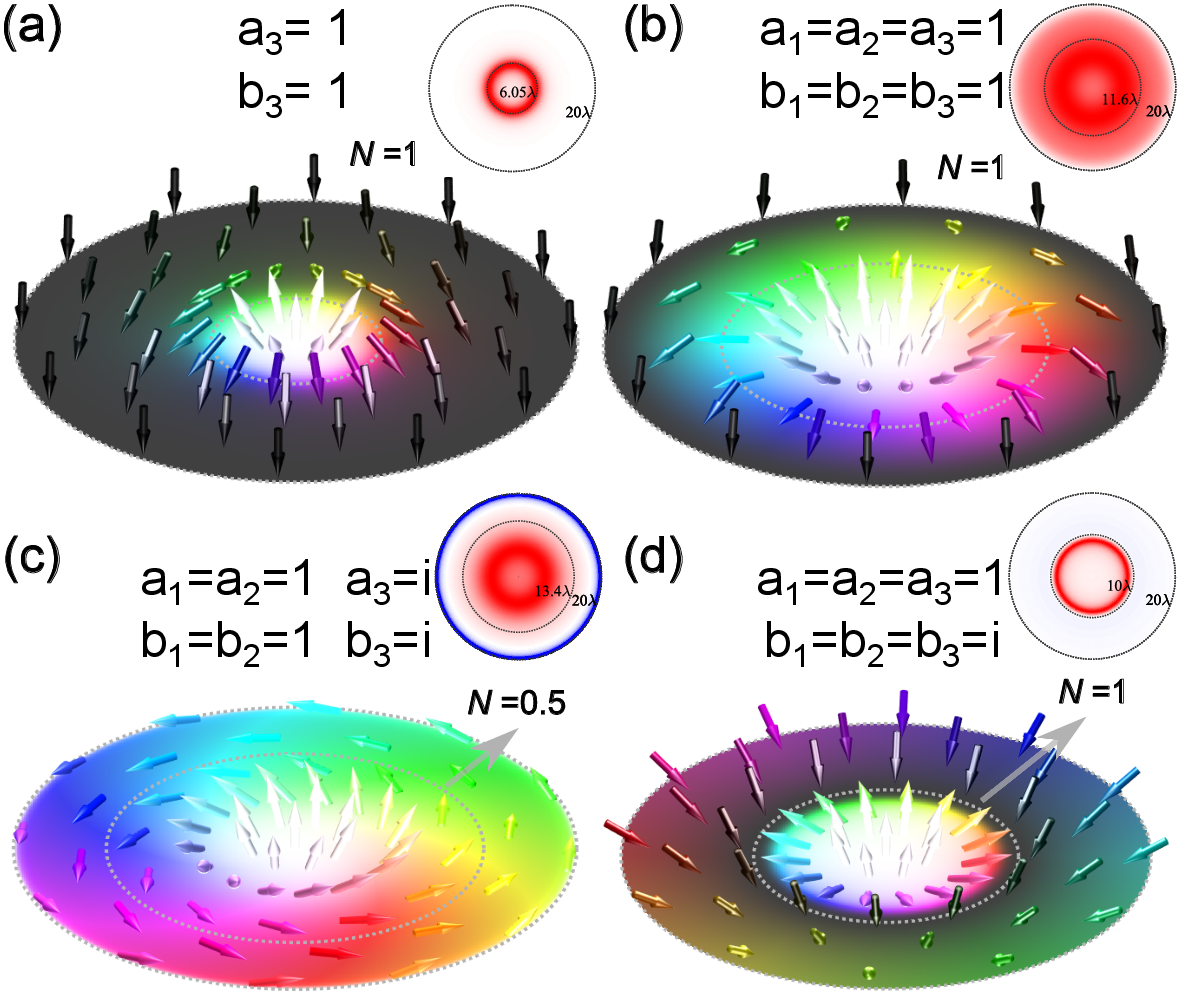}
    \caption{\label{Fig.4} \textbf{Spin angular momentum textures of scattered fields} when the sources are \textbf{(a)} pure octupoles ($a_3=b_3=1$), \textbf{(b)} mixed multipoles ($a_{1,2,3}=b_{1,2,3}=1$), \textbf{(c)} mixed multipoles with phase difference between different orders ($a_{1,2}=b_{1,2}=1 \quad a_3=b_3=i$) and \textbf{(d)} mixed multipoles with phase difference between electric and magnetic components ($a_{1,2,3}=1 \quad b_{1,2,3}=i$). }
\end{figure}

In the following, we elucidate how helicity is influencing angular momentum textures, which is topic garnering significant research interest recently. The nontrivial SAM distribution in our system originates from SAM of incident light. However, the nontrivial SAM distribution is not necessarily related to phase differences of Mie coefficient. When the incident light is circularly polarized $\mathbf{E_{inc}}=E_0 e^{ikz}(\mathbf{\hat{x}} + i\mathbf{\hat{y}})$, the kinetic momentum and canonical momentum texture exhibits no helicity (FIG. \ref{Fig.3}(f)) for the multipole source $a_3=b_3=1$ and $a_{1,2,3}=b_{1,2,3}=1$, respectively. But the corresponding SAM textures are also nontrivial(FIG. \ref{Fig.4}(a)(b)). Besides, SAM topological texture is same to $\mathbf{P}$ topological texture for above three situations, which can also be proven by FIG. \ref{Fig.5}(b). This phenomenon is common in most of situations. But when we introduce a phase difference between electric and magnetic components of multipole sources, such as $a_{1,2,3}=1, b_{1,2,3}=i$, the $\mathbf{S}$ texture and $\mathbf{P}$ texture are different. As shown in FIG. \ref{Fig.4}(d), a skyrmion is formed within $r=10\lambda$ in examined plane, which is different to FIG. \ref{Fig.3}(d).

The similarity between Poynting vector texture and SAM texture of multipole scattering field also provides a potential tool for detecting SAM field, enabling experimental determination of topological quasiparticles. This technique, which uses nanoparticles as probes based on their scattering properties, has been applied in several experiments \cite{eismann2021transverse,banzer2010experimental}.

\section{Topological Stability}

\begin{figure}[h]
    \includegraphics[width=0.49\textwidth]{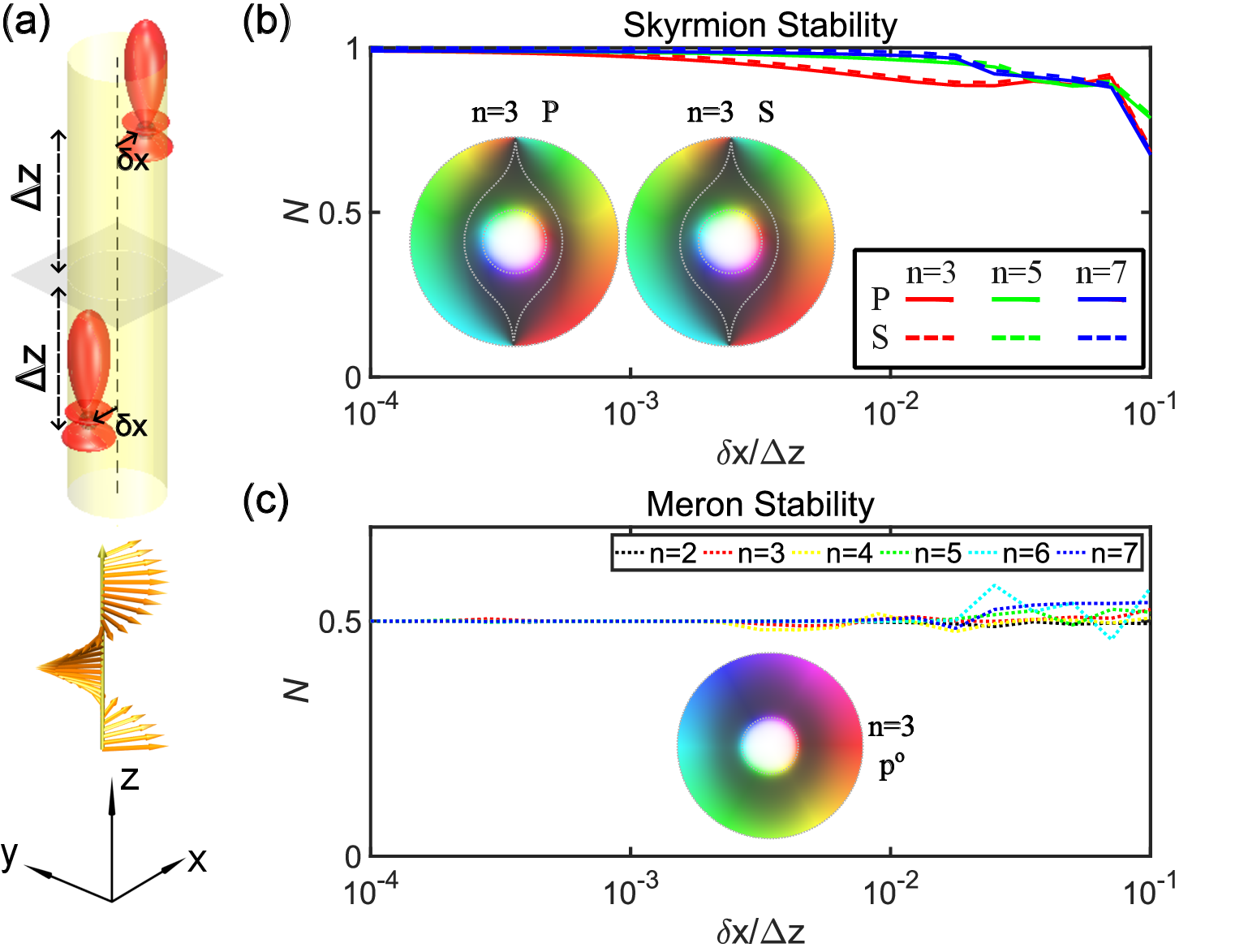}
    \caption{\label{Fig.5} \textbf{Stability of skyrmion and meron.} \textbf{(a)} Schematic diagram of two-particle system with position disturbance. \textbf{(b)} The change of topological charge number in $\mathbf{P}$(full line) and $\mathbf{S}$(broken line) field with respect to $\delta x$. The illustrations indicate the textures when $\delta x=0.01\Delta z$. $n$ is the Mie coefficent order. \textbf{(c)} The change of topological charge number in $\mathbf{p^o}$ field with respect to $\delta x$. And the illustrations indicate the textures when $\delta x=0.01\Delta z$.}
\end{figure}

Topological protection provides stability to these skyrmionic structures. The topological invariant $N$ is expected to remain unperturbed under smooth continuous deformations of the physical domain. We demonstrate this by a transverse translation of the multipoles (FIG. \ref{Fig.5}). 
We compare field textures of $\mathbf{P}$, $\mathbf{S}$ and $\mathbf{p^o}$ for pure multipoles $(a_n=b_n=1)$ . We investigate the change in the topological charge number $N$ of $\mathbf{P}$ and $\mathbf{S}$ with respective $\delta x$, as shown in FIG. \ref{Fig.3}(b). The $\mathbf{P}$ and $\mathbf{S}$ skyrmions we obtained exhibit a certain degree of topological stability. Additionally, as mentioned before, $\mathbf{P}$ and $\mathbf{S}$ topological charge numbers are the same. Higher multipole orders $n$  strengthen the  topological stability of the skyrmion. Also, textures of $\mathbf{P}$ and $\mathbf{S}$ when $a_3=b_3=1$ and $\delta x=0.01\Delta z$ are shown in the illustrations. It's clear that the boundaries of skyrmions are distorted to a spindle curve. Besides, $N$ of $\mathbf{p^o}$ with respective $\delta x$ is shown in FIG. \ref{Fig.5}(c). Compared to $\mathbf{P}$ and $\mathbf{S}$ skyrmion, $\mathbf{p^o}$ merons are more stable. As shown in the illustration, the boundary (inner circle) of $\mathbf{p^o}$ meron is more stable. Besides, for $\mathbf{P}$ and $\mathbf{S}$ texture, the inner merons within the inner circle in illustrations of FIG. \ref{Fig.5}(b) are also stable. These features of topological stability can be explained by directivity of multipole scattering, which is a significant property of multipole. The non-trivial topological structure of $\mathbf{P}$, $\mathbf{S}$ and $\mathbf{p^o}$ were formed based on the directivity, especially the strong forward scattering. For multipoles, the higher the order is, the stronger the forward scattering is, which can explain $\mathbf{P}$ and $\mathbf{S}$ skyrmion is more stable when the sources' orders get higher. In addition, the meron at center is based on the strongest exact forward scattering at center, where the boundary of skyrmion depends on first-order scattering outside the center, which is smaller compared to excat forward scattering. So the merons are more stable than skyrmions here.

\section{Discussion}

In summary, we found a new way to construct topological structures in Mie scattering fields, namely skyrmions in the Poynting vectors $\mathbf{P}$ and merons in the canonical momentum field $\mathbf{p^o}$. Furthermore, our system requires only a single incident beam, making use of the asymmetry of multipole scattered field. In our system, $\mathbf{p^o}$ merons are more stable than $\mathbf{P}$ and $\mathbf{S}$ skyrmions when the scattering particles move away from the optical axis. Higher-order multipole sources result in more stable skyrmions as the multipole sources shift. For theoretical analysis, similar multipole expansions are frequently used in the study of electromagnetic fields, where the fields at distant points are given in terms of sources in a small region. Our research elucidates topological structure formationof multipole scattering. Moreover, multipoles are not limited to Mie scattering systems; they can be observed in other systems, e.g., metasurfaces\cite{babicheva2021multipole,terekhov2019multipole,allayarov2024multiresonant}. These topological structures exhibit good topological stability. And certain types $\mathbf{P}$ or $\mathbf{p^o}$ topological structure may contribute to research on optical forces.


\bibliographystyle{apsrev4-2}

\bibliography{apssamp}

\providecommand{\noopsort}[1]{}\providecommand{\singleletter}[1]{#1}%
\begin{thebibliography}{46}%
\makeatletter
\providecommand \@ifxundefined [1]{%
 \@ifx{#1\undefined}
}%
\providecommand \@ifnum [1]{%
 \ifnum #1\expandafter \@firstoftwo
 \else \expandafter \@secondoftwo
 \fi
}%
\providecommand \@ifx [1]{%
 \ifx #1\expandafter \@firstoftwo
 \else \expandafter \@secondoftwo
 \fi
}%
\providecommand \natexlab [1]{#1}%
\providecommand \enquote  [1]{``#1''}%
\providecommand \bibnamefont  [1]{#1}%
\providecommand \bibfnamefont [1]{#1}%
\providecommand \citenamefont [1]{#1}%
\providecommand \href@noop [0]{\@secondoftwo}%
\providecommand \href [0]{\begingroup \@sanitize@url \@href}%
\providecommand \@href[1]{\@@startlink{#1}\@@href}%
\providecommand \@@href[1]{\endgroup#1\@@endlink}%
\providecommand \@sanitize@url [0]{\catcode `\\12\catcode `\$12\catcode
  `\&12\catcode `\#12\catcode `\^12\catcode `\_12\catcode `\%12\relax}%
\providecommand \@@startlink[1]{}%
\providecommand \@@endlink[0]{}%
\providecommand \url  [0]{\begingroup\@sanitize@url \@url }%
\providecommand \@url [1]{\endgroup\@href {#1}{\urlprefix }}%
\providecommand \urlprefix  [0]{URL }%
\providecommand \Eprint [0]{\href }%
\providecommand \doibase [0]{https://doi.org/}%
\providecommand \selectlanguage [0]{\@gobble}%
\providecommand \bibinfo  [0]{\@secondoftwo}%
\providecommand \bibfield  [0]{\@secondoftwo}%
\providecommand \translation [1]{[#1]}%
\providecommand \BibitemOpen [0]{}%
\providecommand \bibitemStop [0]{}%
\providecommand \bibitemNoStop [0]{.\EOS\space}%
\providecommand \EOS [0]{\spacefactor3000\relax}%
\providecommand \BibitemShut  [1]{\csname bibitem#1\endcsname}%
\let\auto@bib@innerbib\@empty
\bibitem [{\citenamefont {Skyrme}(1962)}]{skyrme1962unified}%
  \BibitemOpen
  \bibfield  {author} {\bibinfo {author} {\bibfnamefont {T.~H.~R.}\
  \bibnamefont {Skyrme}},\ }\href@noop {} {\bibfield  {journal} {\bibinfo
  {journal} {Nuclear Physics}\ }\textbf {\bibinfo {volume} {31}},\ \bibinfo
  {pages} {556} (\bibinfo {year} {1962})}\BibitemShut {NoStop}%
\bibitem [{\citenamefont {Al~Khawaja}\ and\ \citenamefont
  {Stoof}(2001)}]{al2001skyrmions}%
  \BibitemOpen
  \bibfield  {author} {\bibinfo {author} {\bibfnamefont {U.}~\bibnamefont
  {Al~Khawaja}}\ and\ \bibinfo {author} {\bibfnamefont {H.}~\bibnamefont
  {Stoof}},\ }\href@noop {} {\bibfield  {journal} {\bibinfo  {journal}
  {Nature}\ }\textbf {\bibinfo {volume} {411}},\ \bibinfo {pages} {918}
  (\bibinfo {year} {2001})}\BibitemShut {NoStop}%
\bibitem [{\citenamefont {Ruostekoski}\ and\ \citenamefont
  {Anglin}(2001)}]{ruostekoski2001creating}%
  \BibitemOpen
  \bibfield  {author} {\bibinfo {author} {\bibfnamefont {J.}~\bibnamefont
  {Ruostekoski}}\ and\ \bibinfo {author} {\bibfnamefont {J.}~\bibnamefont
  {Anglin}},\ }\href@noop {} {\bibfield  {journal} {\bibinfo  {journal}
  {Physical review letters}\ }\textbf {\bibinfo {volume} {86}},\ \bibinfo
  {pages} {3934} (\bibinfo {year} {2001})}\BibitemShut {NoStop}%
\bibitem [{\citenamefont {Fukuda}\ and\ \citenamefont
  {{\v{Z}}umer}(2011)}]{fukuda2011quasi}%
  \BibitemOpen
  \bibfield  {author} {\bibinfo {author} {\bibfnamefont {J.-i.}\ \bibnamefont
  {Fukuda}}\ and\ \bibinfo {author} {\bibfnamefont {S.}~\bibnamefont
  {{\v{Z}}umer}},\ }\href@noop {} {\bibfield  {journal} {\bibinfo  {journal}
  {Nature communications}\ }\textbf {\bibinfo {volume} {2}},\ \bibinfo {pages}
  {246} (\bibinfo {year} {2011})}\BibitemShut {NoStop}%
\bibitem [{\citenamefont {Duzgun}\ and\ \citenamefont
  {Nisoli}(2021)}]{duzgun2021skyrmion}%
  \BibitemOpen
  \bibfield  {author} {\bibinfo {author} {\bibfnamefont {A.}~\bibnamefont
  {Duzgun}}\ and\ \bibinfo {author} {\bibfnamefont {C.}~\bibnamefont
  {Nisoli}},\ }\href@noop {} {\bibfield  {journal} {\bibinfo  {journal}
  {Physical Review Letters}\ }\textbf {\bibinfo {volume} {126}},\ \bibinfo
  {pages} {047801} (\bibinfo {year} {2021})}\BibitemShut {NoStop}%
\bibitem [{\citenamefont {Tai}\ \emph {et~al.}(2024)\citenamefont {Tai},
  \citenamefont {Hess}, \citenamefont {Wu},\ and\ \citenamefont
  {Smalyukh}}]{tai2024field}%
  \BibitemOpen
  \bibfield  {author} {\bibinfo {author} {\bibfnamefont {J.-S.~B.}\
  \bibnamefont {Tai}}, \bibinfo {author} {\bibfnamefont {A.~J.}\ \bibnamefont
  {Hess}}, \bibinfo {author} {\bibfnamefont {J.-S.}\ \bibnamefont {Wu}},\ and\
  \bibinfo {author} {\bibfnamefont {I.~I.}\ \bibnamefont {Smalyukh}},\
  }\href@noop {} {\bibfield  {journal} {\bibinfo  {journal} {Science Advances}\
  }\textbf {\bibinfo {volume} {10}},\ \bibinfo {pages} {eadj9373} (\bibinfo
  {year} {2024})}\BibitemShut {NoStop}%
\bibitem [{\citenamefont {Muhlbauer}\ \emph {et~al.}(2009)\citenamefont
  {Muhlbauer}, \citenamefont {Binz}, \citenamefont {Jonietz}, \citenamefont
  {Pfleiderer}, \citenamefont {Rosch}, \citenamefont {Neubauer}, \citenamefont
  {Georgii},\ and\ \citenamefont {Boni}}]{muhlbauer2009skyrmion}%
  \BibitemOpen
  \bibfield  {author} {\bibinfo {author} {\bibfnamefont {S.}~\bibnamefont
  {Muhlbauer}}, \bibinfo {author} {\bibfnamefont {B.}~\bibnamefont {Binz}},
  \bibinfo {author} {\bibfnamefont {F.}~\bibnamefont {Jonietz}}, \bibinfo
  {author} {\bibfnamefont {C.}~\bibnamefont {Pfleiderer}}, \bibinfo {author}
  {\bibfnamefont {A.}~\bibnamefont {Rosch}}, \bibinfo {author} {\bibfnamefont
  {A.}~\bibnamefont {Neubauer}}, \bibinfo {author} {\bibfnamefont
  {R.}~\bibnamefont {Georgii}},\ and\ \bibinfo {author} {\bibfnamefont
  {P.}~\bibnamefont {Boni}},\ }\href@noop {} {\bibfield  {journal} {\bibinfo
  {journal} {Science}\ }\textbf {\bibinfo {volume} {323}},\ \bibinfo {pages}
  {915} (\bibinfo {year} {2009})}\BibitemShut {NoStop}%
\bibitem [{\citenamefont {Nagaosa}\ and\ \citenamefont
  {Tokura}(2013)}]{nagaosa2013topological}%
  \BibitemOpen
  \bibfield  {author} {\bibinfo {author} {\bibfnamefont {N.}~\bibnamefont
  {Nagaosa}}\ and\ \bibinfo {author} {\bibfnamefont {Y.}~\bibnamefont
  {Tokura}},\ }\href@noop {} {\bibfield  {journal} {\bibinfo  {journal} {Nature
  nanotechnology}\ }\textbf {\bibinfo {volume} {8}},\ \bibinfo {pages} {899}
  (\bibinfo {year} {2013})}\BibitemShut {NoStop}%
\bibitem [{\citenamefont {Bogdanov}\ and\ \citenamefont
  {Panagopoulos}(2020)}]{bogdanov2020physical}%
  \BibitemOpen
  \bibfield  {author} {\bibinfo {author} {\bibfnamefont {A.~N.}\ \bibnamefont
  {Bogdanov}}\ and\ \bibinfo {author} {\bibfnamefont {C.}~\bibnamefont
  {Panagopoulos}},\ }\href@noop {} {\bibfield  {journal} {\bibinfo  {journal}
  {Nature Reviews Physics}\ }\textbf {\bibinfo {volume} {2}},\ \bibinfo {pages}
  {492} (\bibinfo {year} {2020})}\BibitemShut {NoStop}%
\bibitem [{\citenamefont {Fert}\ \emph {et~al.}(2017)\citenamefont {Fert},
  \citenamefont {Reyren},\ and\ \citenamefont {Cros}}]{fert2017magnetic}%
  \BibitemOpen
  \bibfield  {author} {\bibinfo {author} {\bibfnamefont {A.}~\bibnamefont
  {Fert}}, \bibinfo {author} {\bibfnamefont {N.}~\bibnamefont {Reyren}},\ and\
  \bibinfo {author} {\bibfnamefont {V.}~\bibnamefont {Cros}},\ }\href@noop {}
  {\bibfield  {journal} {\bibinfo  {journal} {Nature Reviews Materials}\
  }\textbf {\bibinfo {volume} {2}},\ \bibinfo {pages} {1} (\bibinfo {year}
  {2017})}\BibitemShut {NoStop}%
\bibitem [{\citenamefont {Han}\ \emph {et~al.}(2022)\citenamefont {Han},
  \citenamefont {Addiego}, \citenamefont {Prokhorenko}, \citenamefont {Wang},
  \citenamefont {Fu}, \citenamefont {Nahas}, \citenamefont {Yan}, \citenamefont
  {Cai}, \citenamefont {Wei}, \citenamefont {Fang} \emph
  {et~al.}}]{han2022high}%
  \BibitemOpen
  \bibfield  {author} {\bibinfo {author} {\bibfnamefont {L.}~\bibnamefont
  {Han}}, \bibinfo {author} {\bibfnamefont {C.}~\bibnamefont {Addiego}},
  \bibinfo {author} {\bibfnamefont {S.}~\bibnamefont {Prokhorenko}}, \bibinfo
  {author} {\bibfnamefont {M.}~\bibnamefont {Wang}}, \bibinfo {author}
  {\bibfnamefont {H.}~\bibnamefont {Fu}}, \bibinfo {author} {\bibfnamefont
  {Y.}~\bibnamefont {Nahas}}, \bibinfo {author} {\bibfnamefont
  {X.}~\bibnamefont {Yan}}, \bibinfo {author} {\bibfnamefont {S.}~\bibnamefont
  {Cai}}, \bibinfo {author} {\bibfnamefont {T.}~\bibnamefont {Wei}}, \bibinfo
  {author} {\bibfnamefont {Y.}~\bibnamefont {Fang}}, \emph {et~al.},\
  }\href@noop {} {\bibfield  {journal} {\bibinfo  {journal} {Nature}\ }\textbf
  {\bibinfo {volume} {603}},\ \bibinfo {pages} {63} (\bibinfo {year}
  {2022})}\BibitemShut {NoStop}%
\bibitem [{\citenamefont {Chen}\ \emph {et~al.}(2024)\citenamefont {Chen},
  \citenamefont {Lourembam}, \citenamefont {Ho}, \citenamefont {Toh},
  \citenamefont {Huang}, \citenamefont {Chen}, \citenamefont {Tan},
  \citenamefont {Yap}, \citenamefont {Lim}, \citenamefont {Tan} \emph
  {et~al.}}]{chen2024all}%
  \BibitemOpen
  \bibfield  {author} {\bibinfo {author} {\bibfnamefont {S.}~\bibnamefont
  {Chen}}, \bibinfo {author} {\bibfnamefont {J.}~\bibnamefont {Lourembam}},
  \bibinfo {author} {\bibfnamefont {P.}~\bibnamefont {Ho}}, \bibinfo {author}
  {\bibfnamefont {A.~K.}\ \bibnamefont {Toh}}, \bibinfo {author} {\bibfnamefont
  {J.}~\bibnamefont {Huang}}, \bibinfo {author} {\bibfnamefont
  {X.}~\bibnamefont {Chen}}, \bibinfo {author} {\bibfnamefont {H.~K.}\
  \bibnamefont {Tan}}, \bibinfo {author} {\bibfnamefont {S.~L.}\ \bibnamefont
  {Yap}}, \bibinfo {author} {\bibfnamefont {R.~J.}\ \bibnamefont {Lim}},
  \bibinfo {author} {\bibfnamefont {H.~R.}\ \bibnamefont {Tan}}, \emph
  {et~al.},\ }\href@noop {} {\bibfield  {journal} {\bibinfo  {journal}
  {Nature}\ }\textbf {\bibinfo {volume} {627}},\ \bibinfo {pages} {522}
  (\bibinfo {year} {2024})}\BibitemShut {NoStop}%
\bibitem [{\citenamefont {Shen}\ \emph
  {et~al.}(2024{\natexlab{a}})\citenamefont {Shen}, \citenamefont {Zhang},
  \citenamefont {Shi}, \citenamefont {Du}, \citenamefont {Yuan},\ and\
  \citenamefont {Zayats}}]{shen2024optical}%
  \BibitemOpen
  \bibfield  {author} {\bibinfo {author} {\bibfnamefont {Y.}~\bibnamefont
  {Shen}}, \bibinfo {author} {\bibfnamefont {Q.}~\bibnamefont {Zhang}},
  \bibinfo {author} {\bibfnamefont {P.}~\bibnamefont {Shi}}, \bibinfo {author}
  {\bibfnamefont {L.}~\bibnamefont {Du}}, \bibinfo {author} {\bibfnamefont
  {X.}~\bibnamefont {Yuan}},\ and\ \bibinfo {author} {\bibfnamefont {A.~V.}\
  \bibnamefont {Zayats}},\ }\href@noop {} {\bibfield  {journal} {\bibinfo
  {journal} {Nature Photonics}\ }\textbf {\bibinfo {volume} {18}},\ \bibinfo
  {pages} {15} (\bibinfo {year} {2024}{\natexlab{a}})}\BibitemShut {NoStop}%
\bibitem [{\citenamefont {Tsesses}\ \emph {et~al.}(2018)\citenamefont
  {Tsesses}, \citenamefont {Ostrovsky}, \citenamefont {Cohen}, \citenamefont
  {Gjonaj}, \citenamefont {Lindner},\ and\ \citenamefont
  {Bartal}}]{tsesses2018optical}%
  \BibitemOpen
  \bibfield  {author} {\bibinfo {author} {\bibfnamefont {S.}~\bibnamefont
  {Tsesses}}, \bibinfo {author} {\bibfnamefont {E.}~\bibnamefont {Ostrovsky}},
  \bibinfo {author} {\bibfnamefont {K.}~\bibnamefont {Cohen}}, \bibinfo
  {author} {\bibfnamefont {B.}~\bibnamefont {Gjonaj}}, \bibinfo {author}
  {\bibfnamefont {N.}~\bibnamefont {Lindner}},\ and\ \bibinfo {author}
  {\bibfnamefont {G.}~\bibnamefont {Bartal}},\ }\href@noop {} {\bibfield
  {journal} {\bibinfo  {journal} {Science}\ }\textbf {\bibinfo {volume}
  {361}},\ \bibinfo {pages} {993} (\bibinfo {year} {2018})}\BibitemShut
  {NoStop}%
\bibitem [{\citenamefont {Davis}\ \emph {et~al.}(2020)\citenamefont {Davis},
  \citenamefont {Janoschka}, \citenamefont {Dreher}, \citenamefont {Frank},
  \citenamefont {zu~Heringdorf},\ and\ \citenamefont
  {Giessen}}]{davis2020ultrafast}%
  \BibitemOpen
  \bibfield  {author} {\bibinfo {author} {\bibfnamefont {T.~J.}\ \bibnamefont
  {Davis}}, \bibinfo {author} {\bibfnamefont {D.}~\bibnamefont {Janoschka}},
  \bibinfo {author} {\bibfnamefont {P.}~\bibnamefont {Dreher}}, \bibinfo
  {author} {\bibfnamefont {B.}~\bibnamefont {Frank}}, \bibinfo {author}
  {\bibfnamefont {F.-J.~M.}\ \bibnamefont {zu~Heringdorf}},\ and\ \bibinfo
  {author} {\bibfnamefont {H.}~\bibnamefont {Giessen}},\ }\href@noop {}
  {\bibfield  {journal} {\bibinfo  {journal} {Science}\ }\textbf {\bibinfo
  {volume} {368}} (\bibinfo {year} {2020})}\BibitemShut {NoStop}%
\bibitem [{\citenamefont {Shen}\ \emph {et~al.}(2021)\citenamefont {Shen},
  \citenamefont {Hou}, \citenamefont {Papasimakis},\ and\ \citenamefont
  {Zheludev}}]{shen2021supertoroidal}%
  \BibitemOpen
  \bibfield  {author} {\bibinfo {author} {\bibfnamefont {Y.}~\bibnamefont
  {Shen}}, \bibinfo {author} {\bibfnamefont {Y.}~\bibnamefont {Hou}}, \bibinfo
  {author} {\bibfnamefont {N.}~\bibnamefont {Papasimakis}},\ and\ \bibinfo
  {author} {\bibfnamefont {N.~I.}\ \bibnamefont {Zheludev}},\ }\href@noop {}
  {\bibfield  {journal} {\bibinfo  {journal} {Nature communications}\ }\textbf
  {\bibinfo {volume} {12}},\ \bibinfo {pages} {5891} (\bibinfo {year}
  {2021})}\BibitemShut {NoStop}%
\bibitem [{\citenamefont {Wang}\ \emph
  {et~al.}(2024{\natexlab{a}})\citenamefont {Wang}, \citenamefont {Bao},
  \citenamefont {Hu}, \citenamefont {Shi}, \citenamefont {Wang}, \citenamefont
  {Zheludev},\ and\ \citenamefont {Shen}}]{wang2024observation}%
  \BibitemOpen
  \bibfield  {author} {\bibinfo {author} {\bibfnamefont {R.}~\bibnamefont
  {Wang}}, \bibinfo {author} {\bibfnamefont {P.-Y.}\ \bibnamefont {Bao}},
  \bibinfo {author} {\bibfnamefont {Z.-Q.}\ \bibnamefont {Hu}}, \bibinfo
  {author} {\bibfnamefont {S.}~\bibnamefont {Shi}}, \bibinfo {author}
  {\bibfnamefont {B.-Z.}\ \bibnamefont {Wang}}, \bibinfo {author}
  {\bibfnamefont {N.~I.}\ \bibnamefont {Zheludev}},\ and\ \bibinfo {author}
  {\bibfnamefont {Y.}~\bibnamefont {Shen}},\ }\href@noop {} {\bibfield
  {journal} {\bibinfo  {journal} {Applied Physics Reviews}\ }\textbf {\bibinfo
  {volume} {11}} (\bibinfo {year} {2024}{\natexlab{a}})}\BibitemShut {NoStop}%
\bibitem [{\citenamefont {Shen}\ \emph
  {et~al.}(2024{\natexlab{b}})\citenamefont {Shen}, \citenamefont
  {Papasimakis},\ and\ \citenamefont {Zheludev}}]{shen2024nondiffracting}%
  \BibitemOpen
  \bibfield  {author} {\bibinfo {author} {\bibfnamefont {Y.}~\bibnamefont
  {Shen}}, \bibinfo {author} {\bibfnamefont {N.}~\bibnamefont {Papasimakis}},\
  and\ \bibinfo {author} {\bibfnamefont {N.~I.}\ \bibnamefont {Zheludev}},\
  }\href@noop {} {\bibfield  {journal} {\bibinfo  {journal} {Nature
  Communications}\ }\textbf {\bibinfo {volume} {15}},\ \bibinfo {pages} {4863}
  (\bibinfo {year} {2024}{\natexlab{b}})}\BibitemShut {NoStop}%
\bibitem [{\citenamefont {Du}\ \emph {et~al.}(2019)\citenamefont {Du},
  \citenamefont {Yang}, \citenamefont {Zayats},\ and\ \citenamefont
  {Yuan}}]{du2019deep}%
  \BibitemOpen
  \bibfield  {author} {\bibinfo {author} {\bibfnamefont {L.}~\bibnamefont
  {Du}}, \bibinfo {author} {\bibfnamefont {A.}~\bibnamefont {Yang}}, \bibinfo
  {author} {\bibfnamefont {A.~V.}\ \bibnamefont {Zayats}},\ and\ \bibinfo
  {author} {\bibfnamefont {X.}~\bibnamefont {Yuan}},\ }\href@noop {} {\bibfield
   {journal} {\bibinfo  {journal} {Nature Physics}\ }\textbf {\bibinfo {volume}
  {15}},\ \bibinfo {pages} {650} (\bibinfo {year} {2019})}\BibitemShut
  {NoStop}%
\bibitem [{\citenamefont {Dai}\ \emph {et~al.}(2020)\citenamefont {Dai},
  \citenamefont {Zhou}, \citenamefont {Ghosh}, \citenamefont {Mong},
  \citenamefont {Kubo}, \citenamefont {Huang},\ and\ \citenamefont
  {Petek}}]{dai2020plasmonic}%
  \BibitemOpen
  \bibfield  {author} {\bibinfo {author} {\bibfnamefont {Y.}~\bibnamefont
  {Dai}}, \bibinfo {author} {\bibfnamefont {Z.}~\bibnamefont {Zhou}}, \bibinfo
  {author} {\bibfnamefont {A.}~\bibnamefont {Ghosh}}, \bibinfo {author}
  {\bibfnamefont {R.~S.}\ \bibnamefont {Mong}}, \bibinfo {author}
  {\bibfnamefont {A.}~\bibnamefont {Kubo}}, \bibinfo {author} {\bibfnamefont
  {C.-B.}\ \bibnamefont {Huang}},\ and\ \bibinfo {author} {\bibfnamefont
  {H.}~\bibnamefont {Petek}},\ }\href@noop {} {\bibfield  {journal} {\bibinfo
  {journal} {Nature}\ }\textbf {\bibinfo {volume} {588}},\ \bibinfo {pages}
  {616} (\bibinfo {year} {2020})}\BibitemShut {NoStop}%
\bibitem [{\citenamefont {Guo}\ \emph {et~al.}(2020)\citenamefont {Guo},
  \citenamefont {Xiao}, \citenamefont {Guo}, \citenamefont {Yuan},\ and\
  \citenamefont {Fan}}]{guo2020meron}%
  \BibitemOpen
  \bibfield  {author} {\bibinfo {author} {\bibfnamefont {C.}~\bibnamefont
  {Guo}}, \bibinfo {author} {\bibfnamefont {M.}~\bibnamefont {Xiao}}, \bibinfo
  {author} {\bibfnamefont {Y.}~\bibnamefont {Guo}}, \bibinfo {author}
  {\bibfnamefont {L.}~\bibnamefont {Yuan}},\ and\ \bibinfo {author}
  {\bibfnamefont {S.}~\bibnamefont {Fan}},\ }\href@noop {} {\bibfield
  {journal} {\bibinfo  {journal} {Physical review letters}\ }\textbf {\bibinfo
  {volume} {124}},\ \bibinfo {pages} {106103} (\bibinfo {year}
  {2020})}\BibitemShut {NoStop}%
\bibitem [{\citenamefont {Lin}\ \emph {et~al.}(2022)\citenamefont {Lin},
  \citenamefont {Du},\ and\ \citenamefont {Yuan}}]{lin2022photonic}%
  \BibitemOpen
  \bibfield  {author} {\bibinfo {author} {\bibfnamefont {M.}~\bibnamefont
  {Lin}}, \bibinfo {author} {\bibfnamefont {L.}~\bibnamefont {Du}},\ and\
  \bibinfo {author} {\bibfnamefont {X.}~\bibnamefont {Yuan}},\ }\href@noop {}
  {\bibfield  {journal} {\bibinfo  {journal} {IEEE Photonics Journal}\ }\textbf
  {\bibinfo {volume} {15}},\ \bibinfo {pages} {1} (\bibinfo {year}
  {2022})}\BibitemShut {NoStop}%
\bibitem [{\citenamefont {Gao}\ \emph {et~al.}(2020)\citenamefont {Gao},
  \citenamefont {Speirits}, \citenamefont {Castellucci}, \citenamefont
  {Franke-Arnold}, \citenamefont {Barnett},\ and\ \citenamefont
  {G{\"o}tte}}]{gao2020paraxial}%
  \BibitemOpen
  \bibfield  {author} {\bibinfo {author} {\bibfnamefont {S.}~\bibnamefont
  {Gao}}, \bibinfo {author} {\bibfnamefont {F.~C.}\ \bibnamefont {Speirits}},
  \bibinfo {author} {\bibfnamefont {F.}~\bibnamefont {Castellucci}}, \bibinfo
  {author} {\bibfnamefont {S.}~\bibnamefont {Franke-Arnold}}, \bibinfo {author}
  {\bibfnamefont {S.~M.}\ \bibnamefont {Barnett}},\ and\ \bibinfo {author}
  {\bibfnamefont {J.~B.}\ \bibnamefont {G{\"o}tte}},\ }\href@noop {} {\bibfield
   {journal} {\bibinfo  {journal} {Physical Review A}\ }\textbf {\bibinfo
  {volume} {102}},\ \bibinfo {pages} {053513} (\bibinfo {year}
  {2020})}\BibitemShut {NoStop}%
\bibitem [{\citenamefont {Shen}\ \emph {et~al.}(2022)\citenamefont {Shen},
  \citenamefont {Mart{\'\i}nez},\ and\ \citenamefont
  {Rosales-Guzm{\'a}n}}]{shen2021generation}%
  \BibitemOpen
  \bibfield  {author} {\bibinfo {author} {\bibfnamefont {Y.}~\bibnamefont
  {Shen}}, \bibinfo {author} {\bibfnamefont {E.~C.}\ \bibnamefont
  {Mart{\'\i}nez}},\ and\ \bibinfo {author} {\bibfnamefont {C.}~\bibnamefont
  {Rosales-Guzm{\'a}n}},\ }\href@noop {} {\bibfield  {journal} {\bibinfo
  {journal} {ACS Photonics}\ }\textbf {\bibinfo {volume} {9}},\ \bibinfo
  {pages} {296} (\bibinfo {year} {2022})}\BibitemShut {NoStop}%
\bibitem [{\citenamefont {Sugic}\ \emph {et~al.}(2021)\citenamefont {Sugic},
  \citenamefont {Droop}, \citenamefont {Otte}, \citenamefont {Ehrmanntraut},
  \citenamefont {Nori}, \citenamefont {Ruostekoski}, \citenamefont {Denz},\
  and\ \citenamefont {Dennis}}]{sugic2021particle}%
  \BibitemOpen
  \bibfield  {author} {\bibinfo {author} {\bibfnamefont {D.}~\bibnamefont
  {Sugic}}, \bibinfo {author} {\bibfnamefont {R.}~\bibnamefont {Droop}},
  \bibinfo {author} {\bibfnamefont {E.}~\bibnamefont {Otte}}, \bibinfo {author}
  {\bibfnamefont {D.}~\bibnamefont {Ehrmanntraut}}, \bibinfo {author}
  {\bibfnamefont {F.}~\bibnamefont {Nori}}, \bibinfo {author} {\bibfnamefont
  {J.}~\bibnamefont {Ruostekoski}}, \bibinfo {author} {\bibfnamefont
  {C.}~\bibnamefont {Denz}},\ and\ \bibinfo {author} {\bibfnamefont {M.~R.}\
  \bibnamefont {Dennis}},\ }\href@noop {} {\bibfield  {journal} {\bibinfo
  {journal} {Nature communications}\ }\textbf {\bibinfo {volume} {12}},\
  \bibinfo {pages} {6785} (\bibinfo {year} {2021})}\BibitemShut {NoStop}%
\bibitem [{\citenamefont {Wang}\ \emph
  {et~al.}(2024{\natexlab{b}})\citenamefont {Wang}, \citenamefont {Zhou},
  \citenamefont {Zheng}, \citenamefont {Sun}, \citenamefont {Cao},
  \citenamefont {Song}, \citenamefont {Deng}, \citenamefont {Qin},
  \citenamefont {Cao},\ and\ \citenamefont {Li}}]{wang2024topological}%
  \BibitemOpen
  \bibfield  {author} {\bibinfo {author} {\bibfnamefont {S.}~\bibnamefont
  {Wang}}, \bibinfo {author} {\bibfnamefont {Z.}~\bibnamefont {Zhou}}, \bibinfo
  {author} {\bibfnamefont {Z.}~\bibnamefont {Zheng}}, \bibinfo {author}
  {\bibfnamefont {J.}~\bibnamefont {Sun}}, \bibinfo {author} {\bibfnamefont
  {H.}~\bibnamefont {Cao}}, \bibinfo {author} {\bibfnamefont {S.}~\bibnamefont
  {Song}}, \bibinfo {author} {\bibfnamefont {Z.-L.}\ \bibnamefont {Deng}},
  \bibinfo {author} {\bibfnamefont {F.}~\bibnamefont {Qin}}, \bibinfo {author}
  {\bibfnamefont {Y.}~\bibnamefont {Cao}},\ and\ \bibinfo {author}
  {\bibfnamefont {X.}~\bibnamefont {Li}},\ }\href@noop {} {\bibfield  {journal}
  {\bibinfo  {journal} {Physical Review Letters}\ }\textbf {\bibinfo {volume}
  {133}},\ \bibinfo {pages} {073802} (\bibinfo {year}
  {2024}{\natexlab{b}})}\BibitemShut {NoStop}%
\bibitem [{\citenamefont {Stratton}(2015)}]{Stratton_2015}%
  \BibitemOpen
  \bibfield  {author} {\bibinfo {author} {\bibfnamefont {J.~A.}\ \bibnamefont
  {Stratton}},\ }\href@noop {} {\emph {\bibinfo {title} {Electromagnetic
  theory}}}\ (\bibinfo  {publisher} {John Wiley \& Sons, Inc},\ \bibinfo {year}
  {2015})\BibitemShut {NoStop}%
\bibitem [{\citenamefont {Poynting}(1884)}]{poynting1884xv}%
  \BibitemOpen
  \bibfield  {author} {\bibinfo {author} {\bibfnamefont {J.~H.}\ \bibnamefont
  {Poynting}},\ }\href@noop {} {\bibfield  {journal} {\bibinfo  {journal}
  {Philosophical Transactions of the Royal Society of London}\ ,\ \bibinfo
  {pages} {343}} (\bibinfo {year} {1884})}\BibitemShut {NoStop}%
\bibitem [{\citenamefont {Gough}(1982)}]{gough1982poynting}%
  \BibitemOpen
  \bibfield  {author} {\bibinfo {author} {\bibfnamefont {W.}~\bibnamefont
  {Gough}},\ }\href@noop {} {\bibfield  {journal} {\bibinfo  {journal}
  {European Journal of Physics}\ }\textbf {\bibinfo {volume} {3}},\ \bibinfo
  {pages} {83} (\bibinfo {year} {1982})}\BibitemShut {NoStop}%
\bibitem [{\citenamefont {Minkowski}(1910)}]{minkowski1910grundgleichungen}%
  \BibitemOpen
  \bibfield  {author} {\bibinfo {author} {\bibfnamefont {H.}~\bibnamefont
  {Minkowski}},\ }\href@noop {} {\bibfield  {journal} {\bibinfo  {journal}
  {Mathematische Annalen}\ }\textbf {\bibinfo {volume} {68}},\ \bibinfo {pages}
  {472} (\bibinfo {year} {1910})}\BibitemShut {NoStop}%
\bibitem [{\citenamefont {Ibrahim}(1909)}]{ibrahim1909elektrodynamik}%
  \BibitemOpen
  \bibfield  {author} {\bibinfo {author} {\bibfnamefont {I.}~\bibnamefont
  {Ibrahim}},\ }\href@noop {} {\bibfield  {journal} {\bibinfo  {journal}
  {Annalen der Physik}\ }\textbf {\bibinfo {volume} {332}},\ \bibinfo {pages}
  {891} (\bibinfo {year} {1909})}\BibitemShut {NoStop}%
\bibitem [{\citenamefont {Abraham}(1910)}]{abraham1910minkowski}%
  \BibitemOpen
  \bibfield  {author} {\bibinfo {author} {\bibfnamefont {M.}~\bibnamefont
  {Abraham}},\ }\href@noop {} {\bibfield  {journal} {\bibinfo  {journal} {Rend.
  Circ. Mat. Palermo}\ }\textbf {\bibinfo {volume} {30}},\ \bibinfo {pages}
  {33} (\bibinfo {year} {1910})}\BibitemShut {NoStop}%
\bibitem [{\citenamefont {Ghosh}\ \emph {et~al.}(2024)\citenamefont {Ghosh},
  \citenamefont {Daniel}, \citenamefont {Gorzkowski}, \citenamefont {Bekshaev},
  \citenamefont {Lapkiewicz},\ and\ \citenamefont
  {Bliokh}}]{ghosh2024canonical}%
  \BibitemOpen
  \bibfield  {author} {\bibinfo {author} {\bibfnamefont {B.}~\bibnamefont
  {Ghosh}}, \bibinfo {author} {\bibfnamefont {A.}~\bibnamefont {Daniel}},
  \bibinfo {author} {\bibfnamefont {B.}~\bibnamefont {Gorzkowski}}, \bibinfo
  {author} {\bibfnamefont {A.~Y.}\ \bibnamefont {Bekshaev}}, \bibinfo {author}
  {\bibfnamefont {R.}~\bibnamefont {Lapkiewicz}},\ and\ \bibinfo {author}
  {\bibfnamefont {K.~Y.}\ \bibnamefont {Bliokh}},\ }\href@noop {} {\bibfield
  {journal} {\bibinfo  {journal} {JOSA B}\ }\textbf {\bibinfo {volume} {41}},\
  \bibinfo {pages} {1276} (\bibinfo {year} {2024})}\BibitemShut {NoStop}%
\bibitem [{\citenamefont {Baxter}\ \emph {et~al.}(1993)\citenamefont {Baxter},
  \citenamefont {Babiker},\ and\ \citenamefont {Loudon}}]{baxter1993canonical}%
  \BibitemOpen
  \bibfield  {author} {\bibinfo {author} {\bibfnamefont {C.}~\bibnamefont
  {Baxter}}, \bibinfo {author} {\bibfnamefont {M.}~\bibnamefont {Babiker}},\
  and\ \bibinfo {author} {\bibfnamefont {R.}~\bibnamefont {Loudon}},\
  }\href@noop {} {\bibfield  {journal} {\bibinfo  {journal} {Physical Review
  A}\ }\textbf {\bibinfo {volume} {47}},\ \bibinfo {pages} {1278} (\bibinfo
  {year} {1993})}\BibitemShut {NoStop}%
\bibitem [{\citenamefont {Afanasev}\ \emph {et~al.}(2022)\citenamefont
  {Afanasev}, \citenamefont {Carlson},\ and\ \citenamefont
  {Mukherjee}}]{afanasev2022superkicks}%
  \BibitemOpen
  \bibfield  {author} {\bibinfo {author} {\bibfnamefont {A.}~\bibnamefont
  {Afanasev}}, \bibinfo {author} {\bibfnamefont {C.~E.}\ \bibnamefont
  {Carlson}},\ and\ \bibinfo {author} {\bibfnamefont {A.}~\bibnamefont
  {Mukherjee}},\ }\href@noop {} {\bibfield  {journal} {\bibinfo  {journal}
  {Physical Review A}\ }\textbf {\bibinfo {volume} {105}},\ \bibinfo {pages}
  {L061503} (\bibinfo {year} {2022})}\BibitemShut {NoStop}%
\bibitem [{\citenamefont {Barnett}\ and\ \citenamefont
  {Berry}(2013)}]{barnett2013superweak}%
  \BibitemOpen
  \bibfield  {author} {\bibinfo {author} {\bibfnamefont {S.~M.}\ \bibnamefont
  {Barnett}}\ and\ \bibinfo {author} {\bibfnamefont {M.}~\bibnamefont
  {Berry}},\ }\href@noop {} {\bibfield  {journal} {\bibinfo  {journal} {Journal
  of Optics}\ }\textbf {\bibinfo {volume} {15}},\ \bibinfo {pages} {125701}
  (\bibinfo {year} {2013})}\BibitemShut {NoStop}%
\bibitem [{\citenamefont {Yang}\ \emph {et~al.}(2022)\citenamefont {Yang},
  \citenamefont {Khosravi},\ and\ \citenamefont {Jacob}}]{yang2022quantum}%
  \BibitemOpen
  \bibfield  {author} {\bibinfo {author} {\bibfnamefont {L.-P.}\ \bibnamefont
  {Yang}}, \bibinfo {author} {\bibfnamefont {F.}~\bibnamefont {Khosravi}},\
  and\ \bibinfo {author} {\bibfnamefont {Z.}~\bibnamefont {Jacob}},\
  }\href@noop {} {\bibfield  {journal} {\bibinfo  {journal} {Physical Review
  Research}\ }\textbf {\bibinfo {volume} {4}},\ \bibinfo {pages} {023165}
  (\bibinfo {year} {2022})}\BibitemShut {NoStop}%
\bibitem [{\citenamefont {Yang}\ \emph {et~al.}(2020)\citenamefont {Yang},
  \citenamefont {Kruk}, \citenamefont {Xu}, \citenamefont {Wang}, \citenamefont
  {Srivastava}, \citenamefont {Koshelev}, \citenamefont {Kravchenko},
  \citenamefont {Singh}, \citenamefont {Han}, \citenamefont {Kivshar} \emph
  {et~al.}}]{yang2020mie}%
  \BibitemOpen
  \bibfield  {author} {\bibinfo {author} {\bibfnamefont {Q.}~\bibnamefont
  {Yang}}, \bibinfo {author} {\bibfnamefont {S.}~\bibnamefont {Kruk}}, \bibinfo
  {author} {\bibfnamefont {Y.}~\bibnamefont {Xu}}, \bibinfo {author}
  {\bibfnamefont {Q.}~\bibnamefont {Wang}}, \bibinfo {author} {\bibfnamefont
  {Y.~K.}\ \bibnamefont {Srivastava}}, \bibinfo {author} {\bibfnamefont
  {K.}~\bibnamefont {Koshelev}}, \bibinfo {author} {\bibfnamefont
  {I.}~\bibnamefont {Kravchenko}}, \bibinfo {author} {\bibfnamefont
  {R.}~\bibnamefont {Singh}}, \bibinfo {author} {\bibfnamefont
  {J.}~\bibnamefont {Han}}, \bibinfo {author} {\bibfnamefont {Y.}~\bibnamefont
  {Kivshar}}, \emph {et~al.},\ }\href@noop {} {\bibfield  {journal} {\bibinfo
  {journal} {Advanced Functional Materials}\ }\textbf {\bibinfo {volume}
  {30}},\ \bibinfo {pages} {1906851} (\bibinfo {year} {2020})}\BibitemShut
  {NoStop}%
\bibitem [{\citenamefont {Sugimoto}\ and\ \citenamefont
  {Fujii}(2021)}]{sugimoto2021colloidal}%
  \BibitemOpen
  \bibfield  {author} {\bibinfo {author} {\bibfnamefont {H.}~\bibnamefont
  {Sugimoto}}\ and\ \bibinfo {author} {\bibfnamefont {M.}~\bibnamefont
  {Fujii}},\ }\href@noop {} {\bibfield  {journal} {\bibinfo  {journal}
  {Advanced Photonics Research}\ }\textbf {\bibinfo {volume} {2}},\ \bibinfo
  {pages} {2000111} (\bibinfo {year} {2021})}\BibitemShut {NoStop}%
\bibitem [{\citenamefont {Allayarov}\ \emph {et~al.}(2024)\citenamefont
  {Allayarov}, \citenamefont {Evlyukhin},\ and\ \citenamefont
  {Cal{\`a}~Lesina}}]{allayarov2024multiresonant}%
  \BibitemOpen
  \bibfield  {author} {\bibinfo {author} {\bibfnamefont {I.}~\bibnamefont
  {Allayarov}}, \bibinfo {author} {\bibfnamefont {A.~B.}\ \bibnamefont
  {Evlyukhin}},\ and\ \bibinfo {author} {\bibfnamefont {A.}~\bibnamefont
  {Cal{\`a}~Lesina}},\ }\href@noop {} {\bibfield  {journal} {\bibinfo
  {journal} {Optics Express}\ }\textbf {\bibinfo {volume} {32}},\ \bibinfo
  {pages} {5641} (\bibinfo {year} {2024})}\BibitemShut {NoStop}%
\bibitem [{\citenamefont {Terekhov}\ \emph {et~al.}(2019)\citenamefont
  {Terekhov}, \citenamefont {Babicheva}, \citenamefont {Baryshnikova},
  \citenamefont {Shalin}, \citenamefont {Karabchevsky},\ and\ \citenamefont
  {Evlyukhin}}]{terekhov2019multipole}%
  \BibitemOpen
  \bibfield  {author} {\bibinfo {author} {\bibfnamefont {P.~D.}\ \bibnamefont
  {Terekhov}}, \bibinfo {author} {\bibfnamefont {V.~E.}\ \bibnamefont
  {Babicheva}}, \bibinfo {author} {\bibfnamefont {K.~V.}\ \bibnamefont
  {Baryshnikova}}, \bibinfo {author} {\bibfnamefont {A.~S.}\ \bibnamefont
  {Shalin}}, \bibinfo {author} {\bibfnamefont {A.}~\bibnamefont
  {Karabchevsky}},\ and\ \bibinfo {author} {\bibfnamefont {A.~B.}\ \bibnamefont
  {Evlyukhin}},\ }\href@noop {} {\bibfield  {journal} {\bibinfo  {journal}
  {Physical Review B}\ }\textbf {\bibinfo {volume} {99}},\ \bibinfo {pages}
  {045424} (\bibinfo {year} {2019})}\BibitemShut {NoStop}%
\bibitem [{\citenamefont {Kerker}\ \emph {et~al.}(1983)\citenamefont {Kerker},
  \citenamefont {Wang},\ and\ \citenamefont
  {Giles}}]{kerker1983electromagnetic}%
  \BibitemOpen
  \bibfield  {author} {\bibinfo {author} {\bibfnamefont {M.}~\bibnamefont
  {Kerker}}, \bibinfo {author} {\bibfnamefont {D.-S.}\ \bibnamefont {Wang}},\
  and\ \bibinfo {author} {\bibfnamefont {C.}~\bibnamefont {Giles}},\
  }\href@noop {} {\bibfield  {journal} {\bibinfo  {journal} {JOSA}\ }\textbf
  {\bibinfo {volume} {73}},\ \bibinfo {pages} {765} (\bibinfo {year}
  {1983})}\BibitemShut {NoStop}%
\bibitem [{\citenamefont {Zhang}\ \emph {et~al.}(2021)\citenamefont {Zhang},
  \citenamefont {Xie}, \citenamefont {Du}, \citenamefont {Shi},\ and\
  \citenamefont {Yuan}}]{zhang2021bloch}%
  \BibitemOpen
  \bibfield  {author} {\bibinfo {author} {\bibfnamefont {Q.}~\bibnamefont
  {Zhang}}, \bibinfo {author} {\bibfnamefont {Z.}~\bibnamefont {Xie}}, \bibinfo
  {author} {\bibfnamefont {L.}~\bibnamefont {Du}}, \bibinfo {author}
  {\bibfnamefont {P.}~\bibnamefont {Shi}},\ and\ \bibinfo {author}
  {\bibfnamefont {X.}~\bibnamefont {Yuan}},\ }\href@noop {} {\bibfield
  {journal} {\bibinfo  {journal} {Physical Review Research}\ }\textbf {\bibinfo
  {volume} {3}},\ \bibinfo {pages} {023109} (\bibinfo {year}
  {2021})}\BibitemShut {NoStop}%
\bibitem [{\citenamefont {Eismann}\ \emph {et~al.}(2021)\citenamefont
  {Eismann}, \citenamefont {Nicholls}, \citenamefont {Roth}, \citenamefont
  {Alonso}, \citenamefont {Banzer}, \citenamefont {Rodr{\'\i}guez-Fortu{\~n}o},
  \citenamefont {Zayats}, \citenamefont {Nori},\ and\ \citenamefont
  {Bliokh}}]{eismann2021transverse}%
  \BibitemOpen
  \bibfield  {author} {\bibinfo {author} {\bibfnamefont {J.}~\bibnamefont
  {Eismann}}, \bibinfo {author} {\bibfnamefont {L.}~\bibnamefont {Nicholls}},
  \bibinfo {author} {\bibfnamefont {D.}~\bibnamefont {Roth}}, \bibinfo {author}
  {\bibfnamefont {M.~A.}\ \bibnamefont {Alonso}}, \bibinfo {author}
  {\bibfnamefont {P.}~\bibnamefont {Banzer}}, \bibinfo {author} {\bibfnamefont
  {F.}~\bibnamefont {Rodr{\'\i}guez-Fortu{\~n}o}}, \bibinfo {author}
  {\bibfnamefont {A.}~\bibnamefont {Zayats}}, \bibinfo {author} {\bibfnamefont
  {F.}~\bibnamefont {Nori}},\ and\ \bibinfo {author} {\bibfnamefont
  {K.}~\bibnamefont {Bliokh}},\ }\href@noop {} {\bibfield  {journal} {\bibinfo
  {journal} {Nature Photonics}\ }\textbf {\bibinfo {volume} {15}},\ \bibinfo
  {pages} {156} (\bibinfo {year} {2021})}\BibitemShut {NoStop}%
\bibitem [{\citenamefont {Banzer}\ \emph {et~al.}(2010)\citenamefont {Banzer},
  \citenamefont {Peschel}, \citenamefont {Quabis},\ and\ \citenamefont
  {Leuchs}}]{banzer2010experimental}%
  \BibitemOpen
  \bibfield  {author} {\bibinfo {author} {\bibfnamefont {P.}~\bibnamefont
  {Banzer}}, \bibinfo {author} {\bibfnamefont {U.}~\bibnamefont {Peschel}},
  \bibinfo {author} {\bibfnamefont {S.}~\bibnamefont {Quabis}},\ and\ \bibinfo
  {author} {\bibfnamefont {G.}~\bibnamefont {Leuchs}},\ }\href@noop {}
  {\bibfield  {journal} {\bibinfo  {journal} {Optics express}\ }\textbf
  {\bibinfo {volume} {18}},\ \bibinfo {pages} {10905} (\bibinfo {year}
  {2010})}\BibitemShut {NoStop}%
\bibitem [{\citenamefont {Babicheva}\ and\ \citenamefont
  {Evlyukhin}(2021)}]{babicheva2021multipole}%
  \BibitemOpen
  \bibfield  {author} {\bibinfo {author} {\bibfnamefont {V.~E.}\ \bibnamefont
  {Babicheva}}\ and\ \bibinfo {author} {\bibfnamefont {A.~B.}\ \bibnamefont
  {Evlyukhin}},\ }\href@noop {} {\bibfield  {journal} {\bibinfo  {journal}
  {Journal of Applied Physics}\ }\textbf {\bibinfo {volume} {129}} (\bibinfo
  {year} {2021})}\BibitemShut {NoStop}%
\end{thebibliography}%

\end{document}